\documentclass[12pt]{article}
\usepackage{graphics}

\def\<~{\stackrel{<}{\mbox{\scriptsize $\sim$}}}
\def\>~{\stackrel{>}{\mbox{\scriptsize $\sim$}}}
\def\eq#1{(\ref{#1})}

\begin{document}

\title{
Cosmological Variation of the Fine Structure Constant from an
Ultra-Light Scalar Field: The Effects of Mass}
\author{
Carl L. Gardner\\
{\em gardner@math.asu.edu}\\
Department of Mathematics and Statistics\\ 
Arizona State University\\
Tempe AZ 85287-1804
}
\date{}

\maketitle
\thispagestyle{empty}

\begin{abstract}
Cosmological variation of the fine structure constant $\alpha$ due to
the evolution of a spatially homogeneous ultra-light scalar field ($m
\sim H_0$) during the matter and $\Lambda$ dominated eras is analyzed.
Agreement of $\Delta \alpha/\alpha$ with the value suggested by recent
observations of quasar absorption lines is obtained by adjusting a
single parameter, the coupling of the scalar field to matter.

Asymptotically $\alpha(t)$ in this model goes to a constant value
$\overline{\alpha} \approx \alpha_0$ in the early radiation and the
late $\Lambda$ dominated eras.  The coupling of the scalar field to
(nonrelativistic) matter drives $\alpha$ slightly away from
$\overline{\alpha}$ in the epochs when the density of matter is
important.

Simultaneous agreement with the more restrictive bounds on the
variation $|\Delta \alpha/\alpha|$ from the Oklo natural fission
reactor and from meteorite samples can be achieved if the mass of the
scalar field is on the order of 0.5--0.6 $H_\Lambda$, where
$H_\Lambda = \Omega_\Lambda^{1/2} H_0$.

Depending on the scalar field mass, $\alpha$ may be slightly smaller
or larger than $\alpha_0$ at the times of big bang nucleosynthesis,
the emission of the cosmic microwave background, the formation of
early solar system meteorites, and the Oklo reactor.  The effects on
the evolution of $\alpha$ due to nonzero mass for the scalar field are
emphasized.

An order of magnitude improvement in the laboratory technique could
lead to a detection of $(\dot{\alpha}/\alpha)_0$.

\end{abstract}

\section{Introduction}

Recent observations by Webb {\em et al.}~\cite{Webb1,Webb2} of
absorption lines in quasar spectra provide evidence for a variation of
the fine structure constant
\begin{equation}
	\frac{\Delta \alpha}{\alpha} = \frac{\alpha(t) - \alpha_0}{\alpha_0}
	= (-0.57 \pm 0.10) \times 10^{-5}
\end{equation}
averaged over the redshift range $0.2 \le z \le 3.7$ (``$\alpha$ was
smaller in the past''), where $\alpha_0$ is the present-day value of
the fine structure constant.  This type of variation of $\alpha$, as
well as variation of other dimensionless coupling constants, is
predicted by theories which unify gravity and the standard model
forces.  For example, string and supergravity theories predict the
existence of massless or ultra-light scalar fields (dilaton or moduli
fields) which through their dynamical evolution can cause temporal
variation of coupling constants.

This investigation will consider cosmological variation of the fine
structure constant due to the evolution of a spatially homogeneous
ultra-light scalar field ($m \sim H_0$, where $H_0$ is the present
value of the Hubble parameter) during the matter and $\Lambda$
dominated eras.  We will assume a flat Friedmann-Robertson-Walker
universe, with $\rho_{c0} = \rho_{m0} + \rho_{r0} + \rho_\Lambda
\approx \rho_{m0} + \rho_\Lambda$ today, where $\rho_{c0}$ is the
present value of the critical density for a flat universe, and
$\rho_{m0}$, $\rho_{r0} \ll \rho_{m0}$ , and $\rho_\Lambda$ are the
present energy densities in (nonrelativistic) matter, radiation, and
the cosmological constant respectively.  Ratios of present energy
densities to the present critical density are denoted by $\Omega_m =
\rho_{m0}/\rho_{c0}$, $\Omega_r = \rho_{r0}/\rho_{c0}$, and
$\Omega_\Lambda = \rho_\Lambda/\rho_{c0}$.

The scalar field $\phi$ may provide the cosmological constant energy
density at the minimum of its potential $V(\overline{\phi})$.  In the
model presented here, the energy density $\rho_\phi -
V(\overline{\phi})$ of the scalar field is always very small compared
with the critical energy density in the radiation, matter, and
$\Lambda$ (dominated) eras, so that the standard
Friedmann-Robertson-Walker evolution of the universe is not affected
by displacements of $\phi$ from $\overline{\phi}$.

Agreement of $\Delta \alpha/\alpha$ with the quasar data can be
obtained by adjusting a single parameter, the coupling of the scalar
field to (nonrelativistic) matter.  Asymptotically $\alpha$ in this
model goes to a constant value $\overline{\alpha} \approx \alpha_0$ in
the early radiation and the late $\Lambda$ eras, insuring agreement
with bounds from cosmic microwave background (CMB) temperature
fluctuations ($|\Delta \alpha/\alpha| < 0.05$ at $z$ = 1090) and
big-bang nucleosynthesis (BBN) ($|\Delta \alpha/\alpha| < 0.02$ at $z
\sim 10^9$--$10^{10}$) (see Ref.~\cite{Uzan} for a comprehensive
review).

Simultaneous agreement with the more restrictive bounds on the total
change $|\Delta \alpha/\alpha| < 10^{-7}$ from $z \approx$ 0.14 to the
present from the Oklo natural fission reactor 1.8 Gyrs
ago~\cite{Oklo1,Oklo2} (by analyzing isotopic ratios of Sm) and
$|\Delta \alpha/\alpha| < 3 \times 10^{-7}$ from $z \approx$ 0.44 to
the present from samples of meteorites formed in the early solar
system 4.6 Gyrs ago~\cite{Olive2} (by analyzing the ratio of
$^{187}$Re to $^{187}$Os) can be achieved if the mass of the scalar
field is on the order of 0.5--0.6 $\Omega_\Lambda^{1/2} H_0$.

The laboratory bounds on the present variation
$|\dot{\alpha}/\alpha|_0 < 3.7 \times 10^{-14}/$yr~\cite{Prestage} and
$(\dot{\alpha}/\alpha)_0 = (-0.4 \pm 16) \times
10^{-16}/$yr~\cite{Sortais} are satisfied for the entire range of
scalar field masses $0 \le m \le 12~\Omega_\Lambda^{1/2} H_0$
considered here.  The variation $(\dot{\alpha}/\alpha)_0$ predicted in
the model may be detectable if the sensitivity of the laboratory
experiments can be increased by an order of magnitude.

\section{The Model}

The scalar field model is based on a generalization of Bekenstein's
model~\cite{Bek} for variable $\alpha$, but with an ultra-light scalar
field mass.  The scalar field obeys the evolution equation
\begin{equation}
	\ddot{\phi} + 3 H \dot{\phi} = -\frac{d V}{d \phi} -
	\zeta_m \frac{\rho_m}{M_*}
\label{phi-eq}
\end{equation}
in the standard Friedmann-Robertson-Walker cosmology.  Here $H$ is the
Hubble parameter, $\zeta_m$ is the coupling of $\phi$ to matter,
$|\zeta_m| \ll 1$, $\rho_m$ is the density of matter, $M_* \<~ M_P $
is the mass scale associated with the scalar field, and the (reduced)
Planck mass $M_P = 2.4 \times 10^{18}$ GeV.  General considerations
show that the coupling of $\phi$ to radiation (including relativistic
matter) should vanish---since $\phi$ couples to the trace of the
energy-momentum tensor for matter and radiation---and that $\zeta_m$
is very nearly constant during the matter and $\Lambda$ eras.

In generalizations~\cite{BSM1}--\cite{Olive} 
of Bekenstein's model, variation of $\alpha$ derives from the coupling
of $\phi$ to the electromagnetic field tensor $F_{\mu \nu}$, through a
term in the action of the form
\begin{equation}
	S_F = \int d^4x \sqrt{-g} \left( - \frac{1}{4} B_F(\phi/M_*) 
	F_{\mu \nu} F^{\mu \nu} \right)
\end{equation}
where $B_F$ is a function (introduced by Damour and
Polyakov~\cite{Damour-Polyakov}) that would be specified by the string
or supergravity theory and constitutes the effective vacuum dielectric
permittivity.  In Bekenstein's model, $B_F$ can be written as $B_F =
\exp\{-2 (\phi - \overline{\phi})/M_*\}$.  Changes in $\phi$ induce
changes in $\alpha$:
\begin{equation}
	\alpha(t) = \frac{\overline{\alpha}}{B_F(\phi(t)/M_*)}
\label{alpha}
\end{equation}
with $B_F(\overline{\phi}/M_*) = 1$.

Our attention will be restricted to small departures of $\phi$ from
$\overline{\phi}$ which will occur in the radiation, matter, and
$\Lambda$ eras.  Defining
\begin{equation}
	\varphi = \frac{\phi - \overline{\phi}}{M_*}
\label{delta}
\end{equation}
the equation for the evolution of the scalar field becomes
\begin{equation}
	\ddot{\varphi} + 3 H \dot{\varphi} + m^2 \varphi =
	- \zeta_m \frac{\rho_m}{M_*^2} =
	- \zeta \frac{\rho_m}{M_P^2} = - \zeta \frac{\rho_{m0}}{M_P^2} 
	\left( \frac{a_0}{a} \right)^3
\label{varphi-eq}
\end{equation}
to first order in $\varphi$, where $m^2 = V''(\overline{\phi})$,
$\zeta = M_P^2 \zeta_m/M_*^2$, $|\zeta| \ll 1$, $a$ is the scale
factor, and $a_0$ is its present value.  For small $\varphi$,
Eq.~\eq{alpha} becomes
\begin{equation}
	\alpha(t) \approx \overline{\alpha} \left( 1 - \zeta_F \varphi \right)
\label{alpha2}
\end{equation}
and
\begin{equation}
	\frac{\Delta \alpha}{\alpha} \approx \zeta_F (\varphi_0 - \varphi)
\end{equation}
where $\zeta_F = B_F'(\overline{\phi}/M_*)$.  In Bekenstein's theory,
$\zeta_F = -2$.

The experimental constraints from the validation of the weak
equivalence principle on the couplings $\zeta$ and $\zeta_F$ may be
evaded by assuming that $\phi$ couples predominantly to dark
matter~\cite{Damour-Gibbons,Olive}.  (For a different view, see the
extended discussion of varying $\alpha$ and the equivalence principle
tests in Ref.~\cite{Damour-equiv}.)

Given a complete particle theory, $\zeta_F$ will be specified and it
will be possible to calculate the coupling $\zeta$ of $\phi$ to
matter.  However, the sign and magnitude of $\zeta$ vary depending on
the way in which Bekenstein's theory is generalized~\cite{Olive}---and
can depend on the unknown properties of dark matter---so here $\zeta$
will simply be determined to fit the quasar data.

One way in which an ultra-light scalar field mass might arise is that
near de Sitter space extrema in four-dimensional extended gauged
supergravity theories (with noncompact internal spaces), there exist
scalar fields with quantized mass squared~\cite{Gates}--\cite{Kallosh}
\begin{equation}
	m^2 = n H_\Lambda^2 ,~~ 
	H_\Lambda^2 =  \frac{\rho_\Lambda}{3 M_P^2} = \Omega_\Lambda H_0^2
\label{m}
\end{equation}
where $-6 \le n \le 12$ is an integer and $H_\Lambda$ is the
asymptotic de Sitter space value of $H$ with cosmological constant
$\rho_\Lambda$.  In certain cases, these theories are directly related
to M/string theory.  An additional advantage of these theories is that
the classical values $m^2 = n H_\Lambda^2$ and $\rho_\Lambda$ are
protected against quantum corrections.  (Cosmological consequences of
such ultra-light scalars in terms of the cosmological constant and the
fate of the universe are discussed in Refs.~\cite{KL1,KL2,KL3}.)

Note that the relation $m^2 = n H_\Lambda^2$ was derived for
supergravity with scalar fields; in the presence of other matter
fields, the relation may be modified.

We will take $n > 0$, corresponding to a de Sitter space minimum, and
will contrast the evolution of $\alpha$ with $n > 0$ with the massless
case $n = 0$.  For $n > 0$, $\varphi \rightarrow 0$ and consequently
$\alpha \rightarrow \overline{\alpha}$ as $t \rightarrow \infty$.  It
is always possible to satisfy the quasar constraints on $\Delta
\alpha/\alpha$ for integer $0 \le n \le 12$, except for $n = 1$.
However, the limits on variation of $\alpha$ from the analyses of Oklo
and meteorite data are not simultaneously satisfied in this model
unless $n$ = 0.24--0.34 ($m \approx$ 0.5--0.6 $H_\Lambda$).

A supergravity inspired potential~\cite{Fre,Kallosh} for the scalar
field is
\begin{equation}
	V(\phi) = \rho_\Lambda \cosh\left(\frac{\sqrt{2} (\phi -
	\overline{\phi})}{M_P}\right) = \rho_\Lambda \cosh(\varphi)
\label{V_ex}
\end{equation}
with $M_* = M_P/\sqrt{2}$.  This potential produces the present-day
cosmological constant $\rho_\Lambda$ when $\phi \approx
\overline{\phi}$ and an ultra-light scalar field mass
\begin{equation}
	m^2(\phi = \overline{\phi}) = \frac{2 \rho_\Lambda}{M_P^2} = 6
	H_\Lambda^2 ~.
\end{equation}
The potential~\eq{V_ex} provides a specific realization of the generic
case~\eq{m}, with $V(\overline{\phi}) = \rho_\Lambda$.  Even in this
case $V(\phi)$ may have a more complicated form in general and only
approach $\rho_\Lambda \cosh(\varphi)$ asymptotically, for example,
after a symmetry breaking phase transition.  (For an analysis of
varying $\alpha$ in models with a ``quintessence'' potential
$V(\phi)$, see Refs.~\cite{quint-1,quint-2}.)

A major difference between the present model and that of
Refs.~\cite{Bek}--\cite{Olive}
is that here the scalar field is assumed to be near the minimum of its
potential, and thus $\alpha(t) \rightarrow \overline{\alpha}$ for $t
\gg H_0^{-1}$.  The initial conditions advocated below also differ
from those of Refs.~\cite{Bek}--\cite{Olive},
and insure that $\alpha$ always remains close to $\overline{\alpha}$.

While a mass term is allowed in the generalized model of Olive and
Pospelov~\cite{Olive}, it is neglected for the explicit solution given
there in Eq.~(3.4).  In a later section, the authors consider a mass
term in the context of the Damour-Polyakov
model~\cite{Damour-Polyakov} for varying constants, where $\zeta_m = 0
= \zeta_F$.  In this model,
\begin{equation}
	\frac{\Delta \alpha}{\alpha} \approx \frac{1}{2} \xi_F 
	\left( \varphi_0^2 - \varphi^2 \right)
\end{equation}
where $\xi_F = B_F''(\overline{\phi}/M_*)$, and the scalar field mass
is given by
\begin{equation}
	m^2 = \frac{\rho_\Lambda}{M_*^2} \left( \xi_\Lambda + 
	\frac{\Omega_m}{\Omega_\Lambda} \xi_m  
	\left( \frac{a_0}{a} \right)^3 \right) ~.
\end{equation}
The authors find that the variation of $\alpha$ can be made {\em
marginally}\/ consistent with the quasar and Oklo data if the
$\xi_\Lambda$ term in $m^2$ is positive and dominant over the $\xi_m$
term.

By contrast, the detailed effects on the evolution of $\alpha$ due to
nonzero mass for the scalar field with $\zeta_F \ne 0$ and $\zeta_m
\ne 0$ will be emphasized below.

\section{Evolution of the Scalar Field}

To determine the initial conditions for the evolution of the scalar
field, we will match approximate solutions to the evolution
equation~\eq{varphi-eq} from the radiation era and the matter era at
the time $t_{m-r}$ ($z \approx$ 3200) of matter-radiation equality.
Note that in the early radiation era, the right-hand side of the
evolution equation~\eq{varphi-eq} goes to zero, since $\zeta_m = 0$
for radiation.

\begin{figure}[htb]
\center{
\scalebox{1.33}{\includegraphics{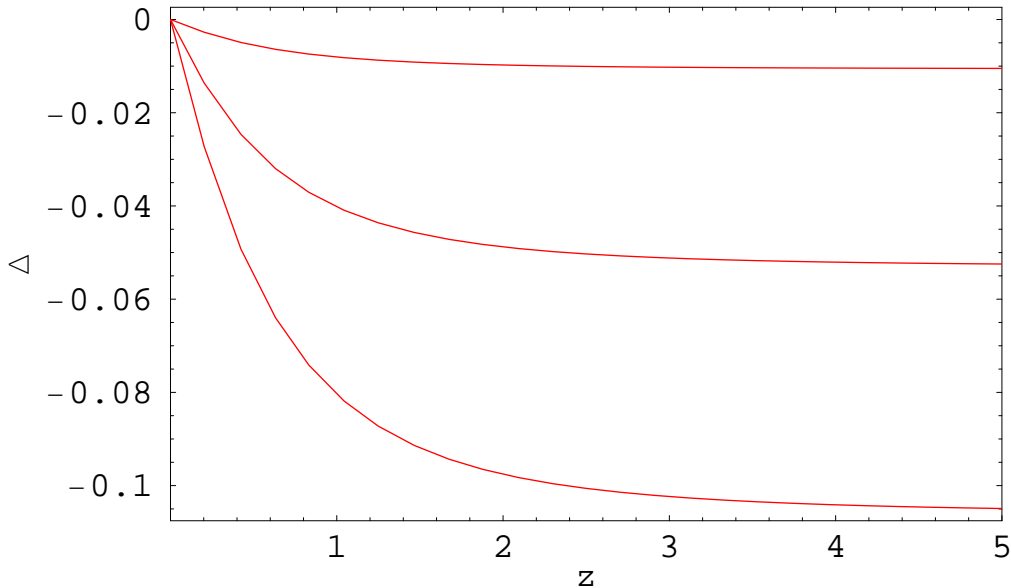}}
}
\caption{$\Delta \equiv \Delta \alpha/\alpha$ vs.\ $z$ for $n = 6$ 
and $\zeta_F = -2$ with various initial conditions 
$\varphi_i = -0.01$, $-0.05$, and $-0.1$, from top to bottom.}
\label{DeltaICs}
\end{figure}

In the early radiation era, $\phi$ may be displaced from
$\overline{\phi}$ and ``frozen'' due to the large frictional term $3 H
\dot{\phi}$ in the evolution equation.  However, for nonzero $m \sim
H_0$, the magnitude of the initial value $|\varphi_i|$ in the early
radiation era must still be $\ll 1$ to satisfy the BBN, CMB, and
quasar bounds on the variation of $\alpha$.  To see this, note that
for $m \sim H_0$, $\varphi$ is frozen near $\varphi_i$ until $H$
becomes of order $H_0$ and then decays with a characteristic timescale
on the order of $H_0^{-1}$.  Thus the change in $\varphi$ from the
early radiation era to the present and the concomitant change $\Delta
\alpha/\alpha$ are on the order of $\varphi_i$ (see
Fig.~\ref{DeltaICs}).  To satisfy the quasar bounds on $\Delta
\alpha/\alpha$ would require fine tuning of initial conditions
($\varphi_i \approx -10^{-5}$).

For $m = 0$, the initial condition for $\varphi$ is irrelevant in the
linearized theory~\eq{varphi-eq} since $\Delta \alpha$ only depends on
changes in $\varphi$; the value of $\varphi_i$ becomes important only
if the scalar potential cannot be neglected.

It is plausible though that $\varphi = 0$ in the early radiation era,
since $V(\phi)$ may have a deep minimum at $\overline{\phi}$ during
inflation in the very early universe, which later, after
one or more phase transitions, becomes shallow with
$V''(\overline{\phi}) \sim H_0$.  For example, in the primordial
inflationary stage, $V''(\overline{\phi})$ may be on the order of
$H_I^2$, where $H_I \gg H_0$ is the Hubble parameter during primordial
inflation.  While the scale factor $a$ inflates by 60 or more
e-foldings, $\phi$ will in this scenario rapidly approach
$\overline{\phi}$.  We therefore take $\varphi = 0 = \dot{\varphi}$ in
the early radiation era.

Equation~\eq{varphi-eq} may be put into dimensionless form by setting
$\tau = H_\Lambda t = \Omega_\Lambda^{1/2} H_0 t$:
\begin{equation}
	\ddot{\varphi} + 3 \frac{H}{H_\Lambda} \dot{\varphi} + n \varphi =
	- 3 \zeta \frac{\Omega_m}{\Omega_\Lambda} 
	\left( \frac{a_0}{a} \right)^3
\label{dimless}
\end{equation}
where henceforth a dot over $\varphi$ denotes differentiation with
respect to $\tau$.

The Hubble parameter is related to the scale factor and the energy
densities in matter, radiation, and the cosmological constant through
the Friedmann equation
\begin{equation}
	\frac{H^2}{H_\Lambda^2} = 
	\left( \frac{1}{a} \frac{d a}{d \tau} \right)^2 =
	\frac{\Omega_m}{\Omega_\Lambda} \left( \frac{a_0}{a} \right)^3 +
	\frac{\Omega_r}{\Omega_\Lambda} \left( \frac{a_0}{a} \right)^4 + 1
\label{H}
\end{equation}
which can be used to solve for $a$ and $H$ in the matter-$\Lambda$,
early matter, and radiation eras.

In the matter-$\Lambda$ era, the scale factor and Hubble parameter
have the explicit forms
\begin{equation}
	\frac{a}{a_0} = \frac{1}{1 + z} =
	\left( \frac{\Omega_m}{\Omega_\Lambda} \right)^{1/3}
	\sinh^{2/3} \left( \frac{3}{2} \tau \right)
\label{a}
\end{equation}
\begin{equation}
	H = H_\Lambda \coth\left(\frac{3}{2} \tau\right)
\label{H-matter-Lambda}
\end{equation}
and the evolution equation becomes
\begin{equation}
	\ddot{\varphi} + 3 \coth\left( \frac{3}{2} \tau \right) \dot{\varphi} 
	+ n \varphi =
	\frac{-3 \zeta}{\sinh^2\left( \frac{3}{2} \tau \right)} ~.
\label{dimless2}
\end{equation}

In the early matter era, the scale factor 
\begin{equation}
	\frac{a}{a_0} \approx 
	\left( \frac{\Omega_m}{\Omega_\Lambda} \right)^{1/3} 
	\left( \frac{3 \tau}{2} \right)^{2/3}
\end{equation}
and the Hubble parameter $H/H_\Lambda \approx 2/(3 \tau)$.  The
mass term $n \varphi$ in Eq.~\eq{dimless} can be neglected in the
early matter (and radiation) eras.  The evolution equation for the
scalar field in the early matter era becomes
\begin{equation}
	\ddot{\varphi} + \frac{2}{\tau} \dot{\varphi} =
	- \frac{4 \zeta}{3 \tau^2} 
\label{varphi-m}
\end{equation}
which has the solution
\begin{equation}
	\varphi_m = - \frac{4 \zeta}{3} \left( \ln \frac{\tau}{\tau_{m-r}} +
	\frac{c_1}{\tau} + c_2
	\right)
\label{sol-m}
\end{equation}
where $\tau_{m-r} = H_\Lambda t_{m-r}$ and $c_1$ and $c_2$ are
constants of integration.

In the radiation era, the scale factor 
\begin{equation}
	\frac{a}{a_0} \approx 
	\left( \frac{4 \Omega_r}{\Omega_\Lambda} \right)^{1/4} 
	s^{1/2} ,~~ s = \tau - \frac{\tau_{m-r}}{4}
\end{equation}
and the Hubble parameter $H/H_\Lambda \approx 1/(2 s)$.  The time
shift $\tau_{m-r}/4$ is determined by matching the Hubble parameters
from the radiation and early matter eras at $\tau_{m-r}$.  (The formal
mathematical singularity at $a \rightarrow 0$ now occurs at $t =
t_{m-r}/4$ due to the choice of the zero of time in Eq.~\eq{a}.)  The
evolution equation for the scalar field becomes
\begin{equation}
	\ddot{\varphi} + \frac{3}{2 s} \dot{\varphi} =
	\frac{-3 \zeta}{4 s_{m-r}^{1/2} s^{3/2}} ~.
\label{varphi-r}
\end{equation}
The solution in the radiation era is
\begin{equation}
	\varphi_r = - \frac{3 \zeta}{2} 
	\left( \frac{s^{1/2}}{s_{m-r}^{1/2}} +
	\frac{c_3}{s^{1/2}} + c_4
	\right)
\label{sol-r}
\end{equation}
where $c_3$ and $c_4$ are constants of integration.

The initial conditions for the scalar field in the early radiation era
are $\varphi_r(\tau_i) = 0 = \dot{\varphi}_r(\tau_i)$, where the
initial time $\tau_i$ satisfies $s_i \ll s_{m-r}$.  These initial
conditions fix the constants in the solution~\eq{sol-r}, yielding
\begin{equation}
	\varphi_r = - \frac{3 \zeta}{2} 
	\left( \frac{s^{1/2}}{s_{m-r}^{1/2}} +
	\frac{s_i}{s_{m-r}^{1/2} s^{1/2}} - \frac{2 s_i^{1/2}}{s_{m-r}^{1/2}}
	\right) ~.
\label{sol-r2}
\end{equation}
Next take the limit $s_i/s_{m-r} \rightarrow 0$ to obtain
\begin{equation}
	\varphi_r \approx - \frac{3 \zeta}{2} 
	\frac{s^{1/2}}{s_{m-r}^{1/2}} ~.
\label{sol-r3}
\end{equation}

Now match $\varphi_r = \varphi_m$ and $\dot{\varphi}_r =
\dot{\varphi}_m$ at $\tau_{m-r}$ to determine the constants in the
solution~\eq{sol-m}:
\begin{equation}
	\varphi_m = - \frac{4 \zeta}{3} \left( \ln \frac{\tau}{\tau_{m-r}}
	+ \frac{\tau_{m-r}}{4 \tau} + \frac{7}{8}
	\right)
\label{sol-m2}
\end{equation}
\begin{equation}
	\dot{\varphi}_m = - \frac{4 \zeta}{3} \left( \frac{1}{\tau}
	- \frac{\tau_{m-r}}{4 \tau^2}
	\right) ~.
\label{velocity-m2}
\end{equation}

To simulate the evolution of the scalar field in the matter-$\Lambda$
era, we use Eq.~\eq{dimless2} with initial conditions provided by
Eqs.~\eq{sol-m2} and~\eq{velocity-m2} evaluated at $\tau_{m-r}$.

\section{Comparison with Quasar, Meteorite, and Oklo Data}

In numerical values for expressions, we take $\Omega_m$ = 0.27,
$\Omega_\Lambda$ = 0.73, and $H_0$ = 71 (km/sec)/Mparsec = $1.5 \times
10^{-33}$ eV, from Table 10 of the first-year WMAP
observations~\cite{WMAP}.  With these parameters, the age of the
universe is $t_0$ = 13.7 Gyrs, and the absorption clouds at $z$ =
0.2--3.7 date to 2.4--11.9 Gyrs ago.

\begin{figure}[t]
\center{
\scalebox{1.33}{\includegraphics{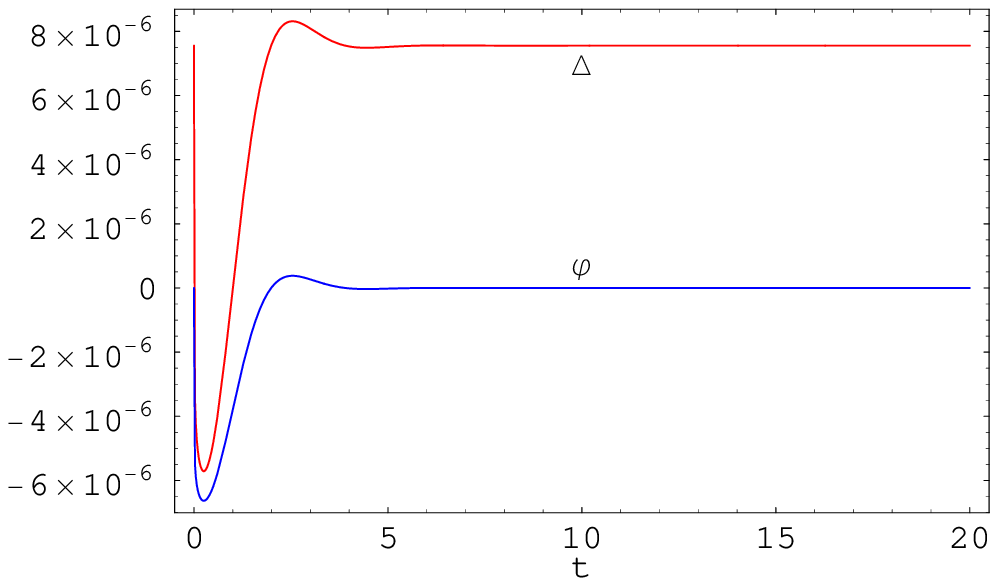}}
}
\caption{Scalar field $\varphi$ and $\Delta \alpha/\alpha$  
vs. $t/t_0$ for $n = 6$.}
\label{phi6}
\end{figure}

\begin{figure}[b]
\center{
\scalebox{1.33}{\includegraphics{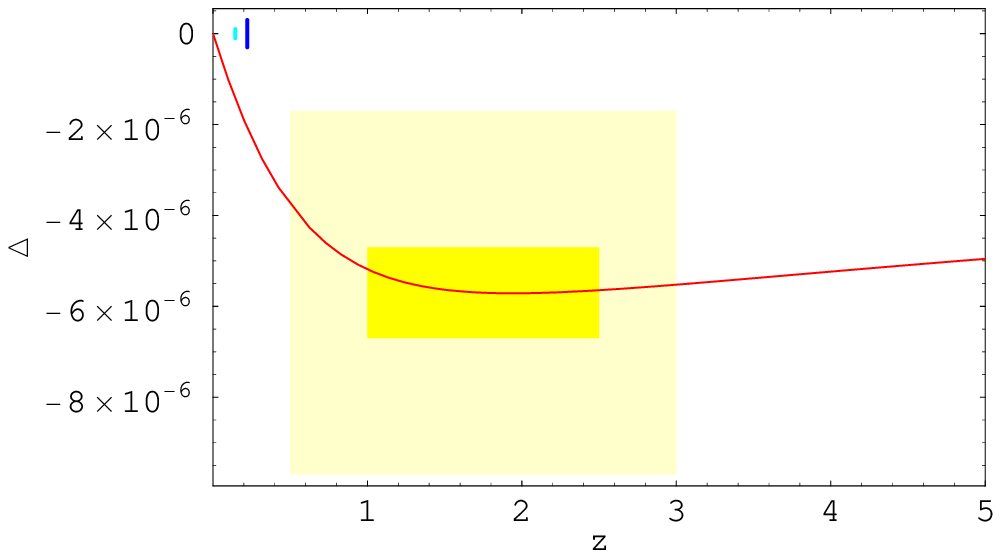}}
}
\caption{$\Delta \alpha/\alpha$ vs.\ $z$ for $n = 6$.}
\label{Delta6}
\end{figure}

\begin{figure}[htbp]
\center{
\scalebox{1.33}{\includegraphics{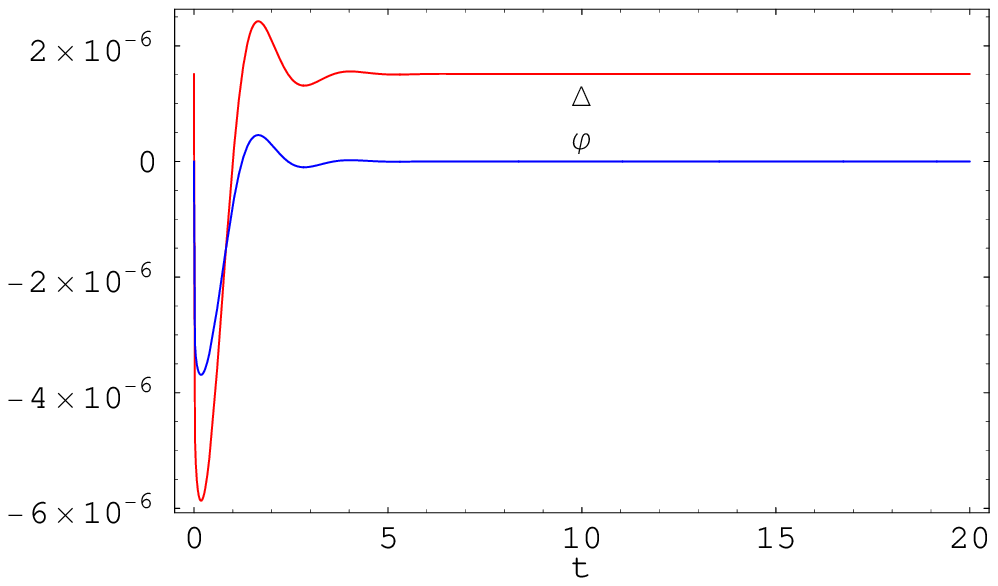}}
}
\caption{Scalar field $\varphi$ and $\Delta \alpha/\alpha$ 
vs. $t/t_0$ for $n = 12$.}
\label{phi12}
\end{figure}

\begin{figure}[htbp]
\center{
\scalebox{1.33}{\includegraphics{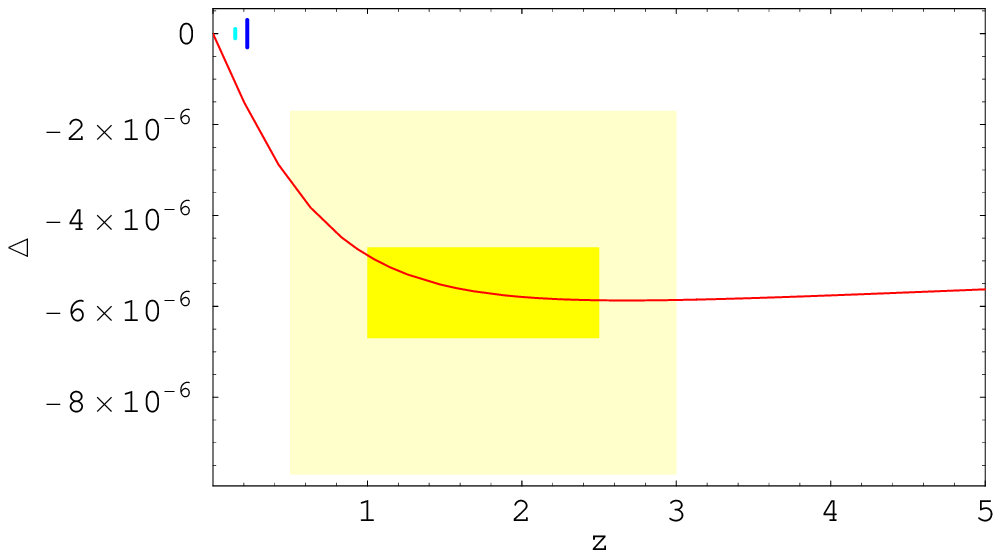}}
}
\caption{$\Delta \alpha/\alpha$ vs.\ $z$ for $n = 12$.}
\label{Delta12}
\end{figure}

\begin{figure}[htbp]
\center{
\scalebox{1.33}{\includegraphics{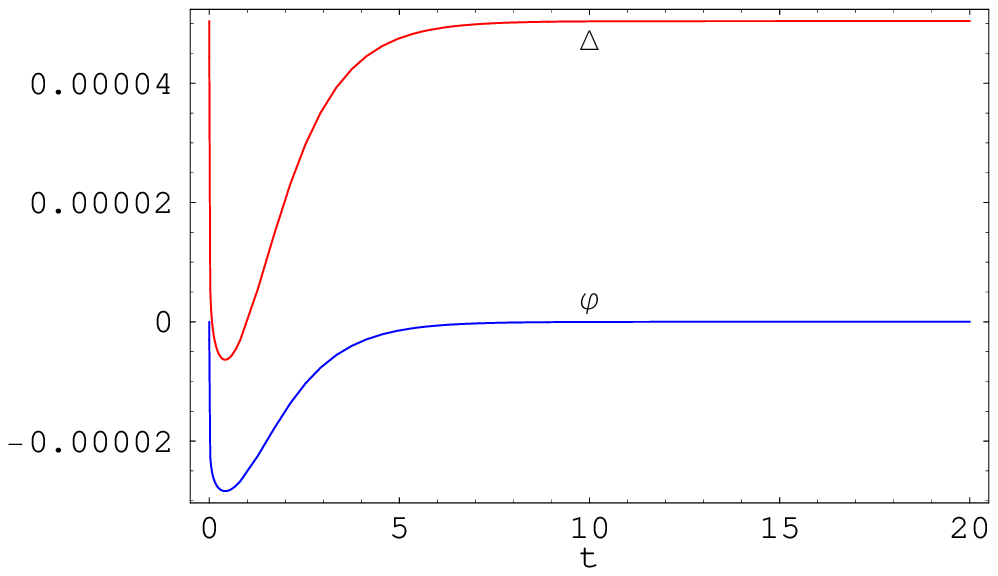}}
}
\caption{Scalar field $\varphi$ and $\Delta \alpha/\alpha$ 
vs. $t/t_0$ for $n = 2$.}
\label{phi2}
\end{figure}

\begin{figure}[htbp]
\center{
\scalebox{1.33}{\includegraphics{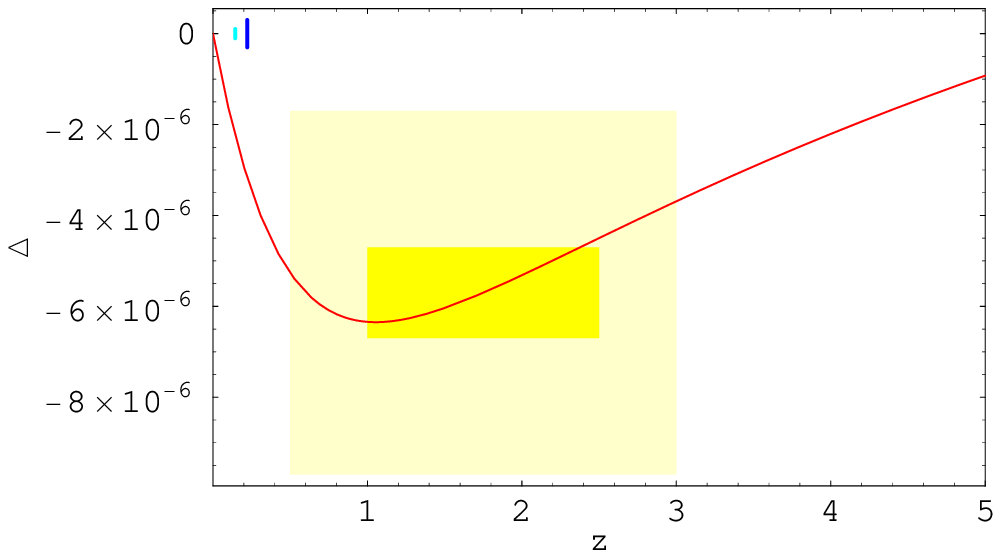}}
}
\caption{$\Delta \alpha/\alpha$ vs.\ $z$ for $n = 2$.}
\label{Delta2}
\end{figure}

\begin{figure}[htbp]
\center{
\scalebox{1.33}{\includegraphics{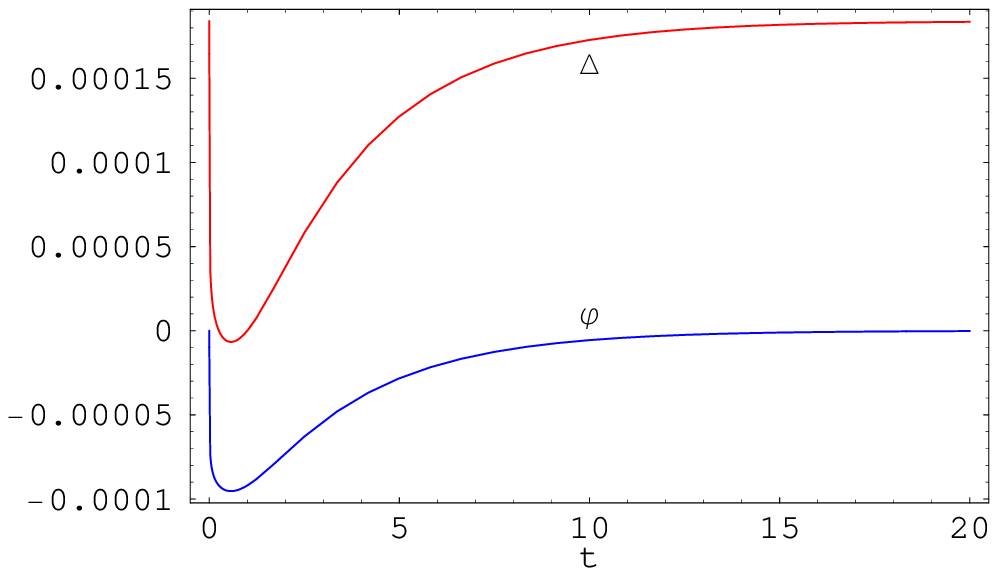}}
}
\caption{Scalar field $\varphi$ and $\Delta \alpha/\alpha$ 
vs. $t/t_0$ for $n = 1$.}
\label{phi1}
\end{figure}

\begin{figure}[htbp]
\center{
\scalebox{1.33}{\includegraphics{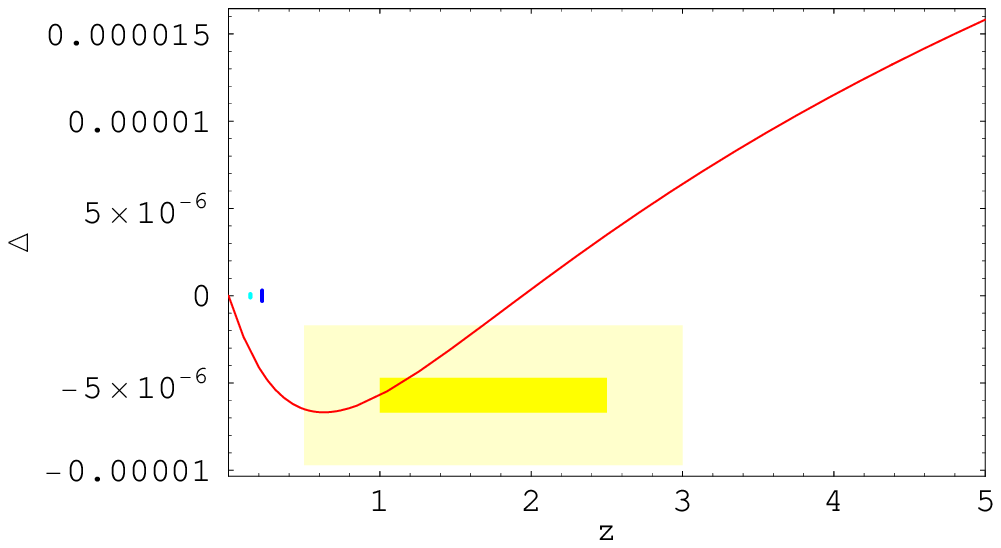}}
}
\caption{$\Delta \alpha/\alpha$ vs.\ $z$ for $n = 1$.}
\label{Delta1}
\end{figure}

\begin{figure}[htbp]
\center{
\scalebox{1.33}{\includegraphics{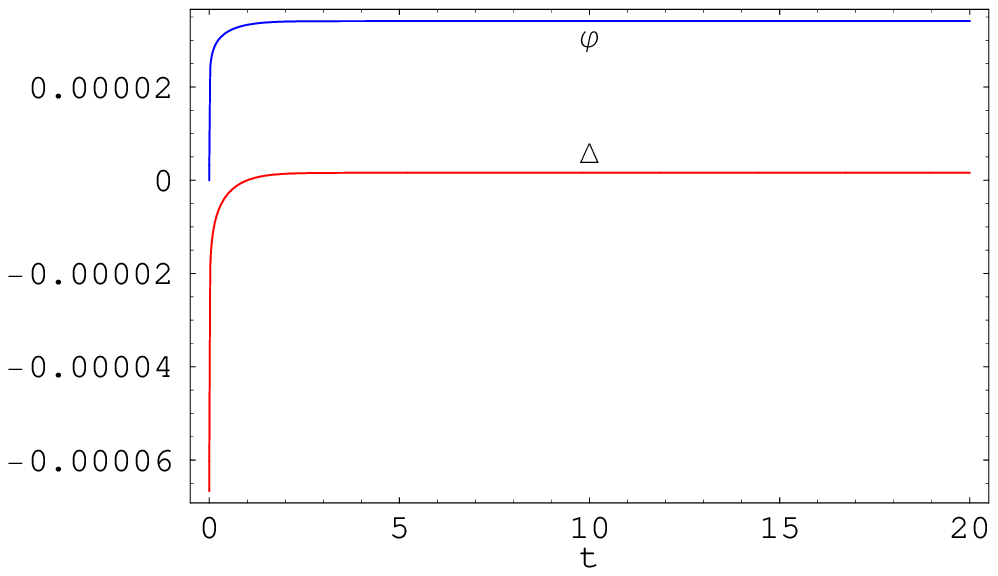}}
}
\caption{Scalar field $\varphi$ and $\Delta \alpha/\alpha$ 
vs. $t/t_0$ for $n = 0$.}
\label{phi0}
\end{figure}

\begin{figure}[htbp]
\center{
\scalebox{1.33}{\includegraphics{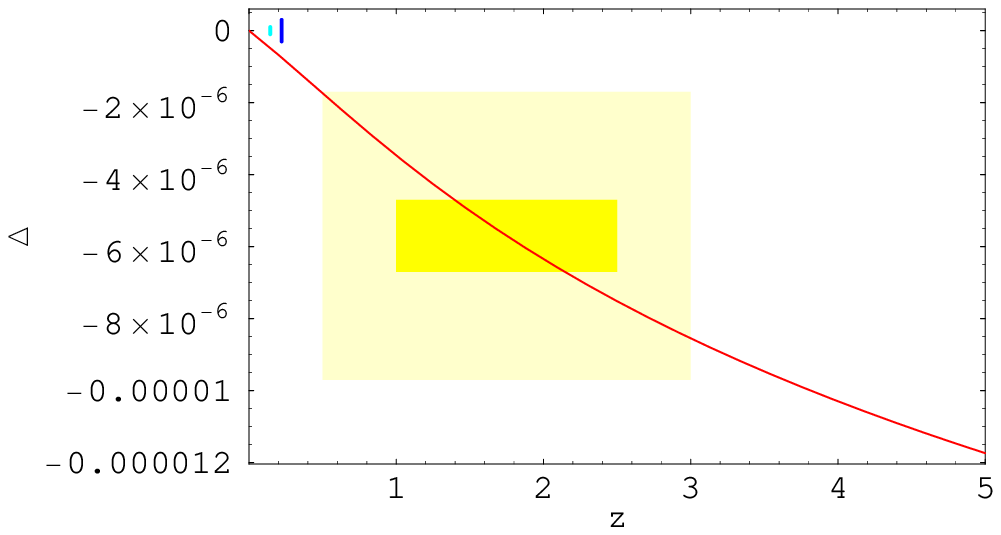}}
}
\caption{$\Delta \alpha/\alpha$ vs.\ $z$ for $n = 0$.}
\label{Delta0}
\end{figure}

\begin{figure}[htbp]
\center{
\scalebox{1.33}{\includegraphics{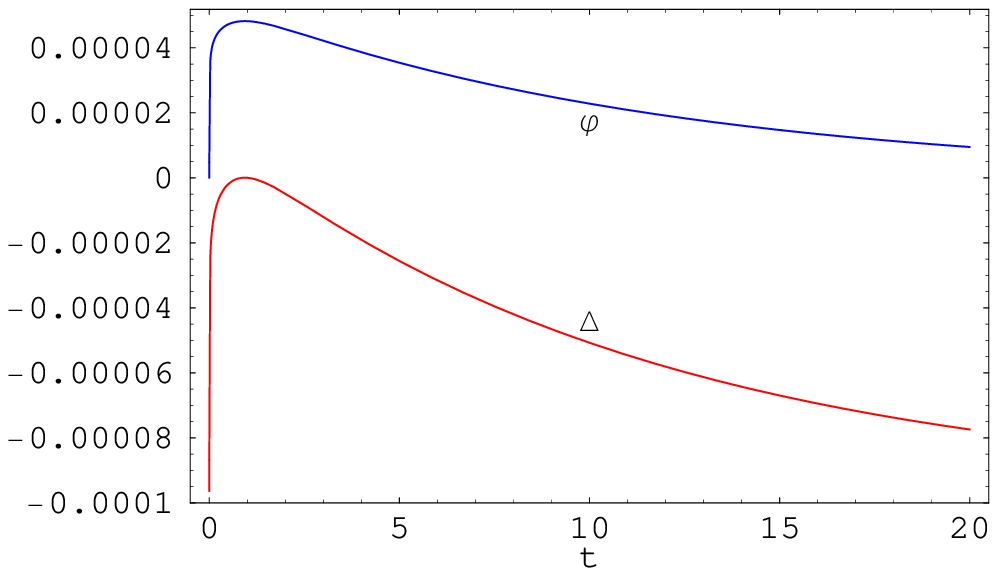}}
}
\caption{Scalar field $\varphi$ and $\Delta \alpha/\alpha$ 
vs. $t/t_0$ for $n = 0.3$.}
\label{phi.3}
\end{figure}

\begin{figure}[htbp]
\center{
\scalebox{1.33}{\includegraphics{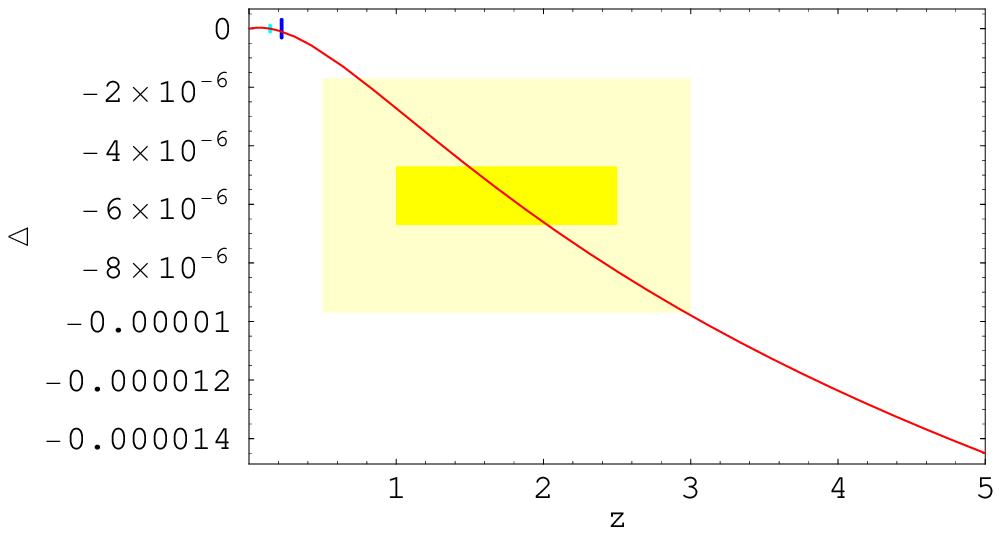}}
}
\caption{$\Delta \alpha/\alpha$ vs.\ $z$ for $n = 0.3$.}
\label{Delta.3}
\end{figure}

Figs.~\ref{phi6}--\ref{Delta.3} present simulations of the evolution
of the scalar field and $\Delta \alpha/\alpha$ for $n$ = 6, 12, 2, 1,
0, and 0.3.

In the figures for $\Delta \alpha(z)/\alpha$, the dark (1 $\sigma$
error bounds for $1 \le z \le 2.5$) and light (a rough guide to the
error bars for $0.6 \le z \le 3$) boxes indicate the quasar bounds
from the bottom panel of Fig.~2 of Ref.~\cite{Webb2}, while the short
vertical lines at $z = 0.14$ and $z = 0.44/2$ indicate the Oklo and
approximate meteorite bounds respectively.  While the Oklo event is
sharply located in time, the ratio of $^{187}$Re to $^{187}$Os
observed today in meteorites involves the total change in $\alpha$
since the time of formation of the meteorites, which is calculated by
averaging $\alpha - \alpha_0$:
\begin{equation}
	\left( \frac{\Delta \alpha}{\alpha} \right)_{meteor} = 
	\frac{1}{t_0 - t} \int_{t}^{t_0} \frac{\alpha(s)}{\alpha_0} ds - 1 ~.
\label{change}
\end{equation}

\begin{figure}[htbp]
\center{
\scalebox{1.33}{\includegraphics{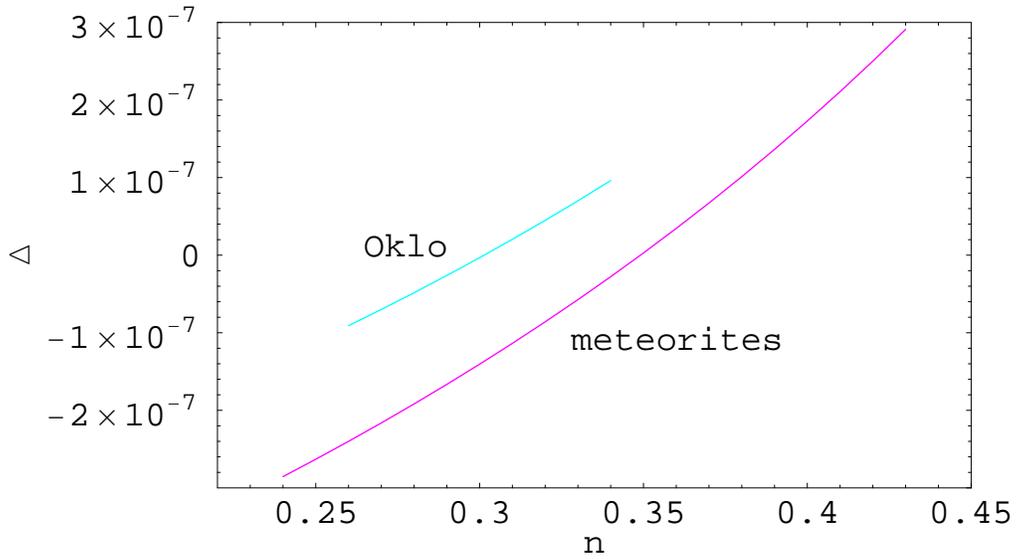}}
}
\caption{$\Delta \alpha/\alpha$ vs.\ $n$ satisfying the Oklo and 
meteorite bounds.}
\label{OkloMets}
\end{figure}

The scalar field $\varphi$ solving the initial value problem defined
in Eqs.~\eq{dimless2}, \eq{sol-m2}, and~\eq{velocity-m2} will be
proportional to $\zeta$, and thus $\Delta \alpha/\alpha$ will be
proportional to $\zeta \zeta_F$.  For simplicity we will set $\zeta_F
= -2$, but a general value for $\zeta_F$ can be reinserted.  For $0
\le n \le 12$ except $n = 1$, $\zeta$ is fixed by setting $\Delta
\alpha/\alpha = -0.57 \times 10^{-5}$ at $z$ = 1.75.  For $n = 1$,
better results were obtained by setting $\Delta \alpha/\alpha = -0.57
\times 10^{-5}$ at $z$ = 1 (see Fig.~\ref{Delta1}).  The massless case
shown in Fig.~\ref{phi0} agrees with Eq.~(3.4) of Ref.~\cite{Olive}
with $\zeta_\Lambda = 0$.  For $n = 0.3$, Figs.~\ref{Delta.3}
and~\ref{OkloMets} show that the quasar, meteorite, and Oklo bounds
can be satisfied simultaneously.

The number of visible oscillations in the scalar field $\varphi$ (and
thus also in $\Delta \alpha/\alpha$) corresponds to how massive the
scalar field is, with at one extreme no oscillations for the massless
case (Fig.~\ref{phi0}), and at the other extreme two visible
oscillations for the $n$ = 12 case (Fig.~\ref{phi12}).

In this model, for $n > 1$ ($n = 1$), $\alpha$ was actually larger in
the past at some point before the period of the $z$ = 0.2--3.7 ($z$ =
0.2--1.9) absorption clouds, and will be larger again in the future.
For $n = 0$, $\alpha$ was smaller in the past, but will be very
slightly larger in the future.  And for $n = 0.3$, $\alpha$ was
smaller in the past until just after $t_{Oklo}$, was slightly larger
from that point up to $t_0$, and then again will be smaller in the
future ($\Delta \alpha/\alpha \rightarrow - 10^{-4}$).

Values of $\zeta$, BBN, CMB, meteorite, and Oklo $\Delta
\alpha/\alpha$, and $(\dot{\alpha}/\alpha)_0$ for various scalar field
masses are presented in Table 1.

The BBN, CMB, and quasar bounds on $\Delta \alpha/\alpha$ and the
laboratory bound on $|\dot{\alpha}/\alpha|_0$ are satisfied for
integer $0 \le n \le 12$, except that the $n = 1$ case cannot be made
to satisfy the quasar bounds in this model.  The variation $|\Delta
\alpha/\alpha|$ satisfies in addition the Oklo bound for $0.26 \le n
\le 0.34$ and the meteorite bound for $0.24 \le n \le 0.43$ (see
Fig.~\ref{OkloMets}).  There is a small range of scalar field masses
for which $|\Delta \alpha/\alpha|_{Oklo} < 10^{-8}$ (and can be made
to go to zero by extreme fine tuning).  For this range of $n$,
$(\Delta \alpha/\alpha)_{meteor} \approx -1 \times 10^{-7}$.

Depending on the scalar field mass, the predicted BBN, CMB, meteorite,
and Oklo values of $\alpha$ may be slightly smaller or larger than
$\alpha_0$.  Note that the sign of $\zeta$ for $n = 0$ is opposite to
the sign for $n \ge 1$.  An order of magnitude improvement in the
experimental technique could lead to a detection of
$(\dot{\alpha}/\alpha)_0$.

\begin{figure}[htbp]
\center{
\scalebox{1.33}{\includegraphics{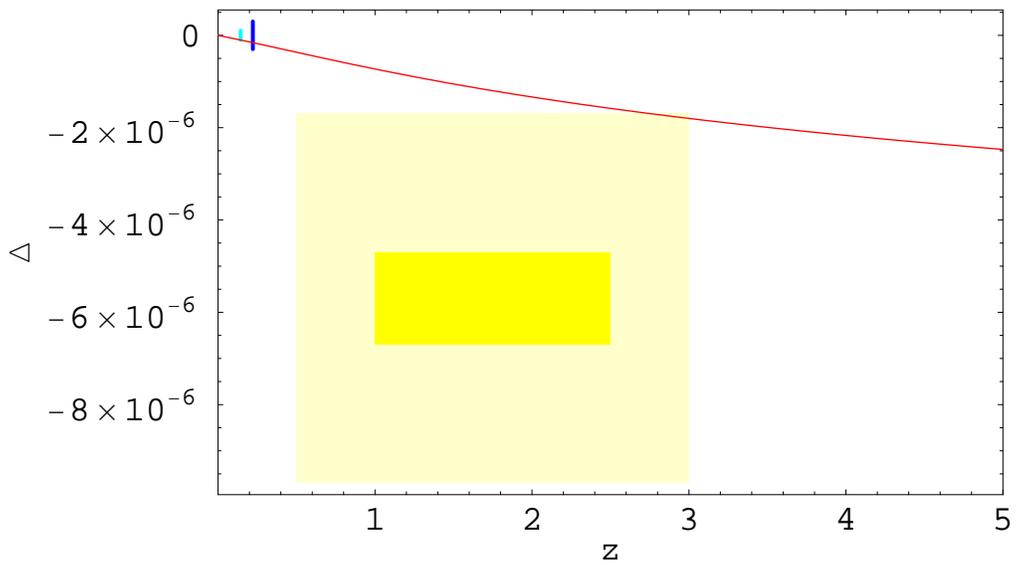}}
}
\caption{$\Delta \alpha/\alpha$ vs.\ $z$ for $n = 0$ by setting
$\Delta \alpha/\alpha = -0.18 \times 10^{-5}$ at $z$ = 3.}
\label{DeltaAdj}
\end{figure}

For the massless case, the variation in $\alpha$ can be made
marginally consistent with the quasar, meteorite, and Oklo bounds by
setting $\Delta \alpha/\alpha = -0.18 \times 10^{-5}$ at $z$ = 3
(Fig.~\ref{DeltaAdj}).

The behavior of $\alpha(z)$ can pin down the values for $\zeta$ and
$m$.  Conversely, even knowing only the sign of $\zeta$ or $\Delta
\alpha/\alpha$ can rule out certain values of the scalar field mass.
For example, if $(\Delta \alpha/\alpha)_{BBN} > 0$ or $\zeta \zeta_F <
0$, then $n \>~ 1$.

\section{Conclusion}

Asymptotically $\alpha(t)$ in this model goes to a constant value
$\overline{\alpha} \approx \alpha_0$ in the early radiation and the
late $\Lambda$ dominated eras.  The coupling of the scalar field to
(nonrelativistic) matter drives $\alpha$ slightly away from
$\overline{\alpha}$ in the epochs when the density of matter is
important.

Even for $\Omega_\Lambda = 0$, $\alpha \rightarrow \overline{\alpha}$
as $t \rightarrow \infty$ as long as $m \ne 0$.  In the massless case,
as $t \rightarrow \infty$, $\alpha$ goes to a constant value which
differs from $\overline{\alpha}$ but still approximately equals 1/137
for $\Omega_\Lambda > 0$, while if $\Omega_\Lambda = 0$ (and $V(\phi)
\equiv 0$), $\Delta \alpha \sim \zeta \zeta_F \ln(\tau/\tau_0)$, as in
Ref.~\cite{BSM2}.  Thus the variation $|\Delta \alpha/\alpha|$ of the
fine structure constant becomes of order 1 only if both $\rho_\Lambda
\rightarrow 0$ and $m = 0$, and only for $t \gg t_0$.

The simulations above indicate that it is possible to extract
properties of the scalar field from quasar absorption line spectra,
including the coupling of $\phi$ to matter and its mass.  The
variation of $\alpha$ has different behaviors in the
redshift range $0 \le z \le 5$ depending on the mass of the scalar
field.  Thus additional quasar absorption line data, and better Oklo
and meteorite bounds, will help elucidate the properties of the scalar
field.  The case $m = H_\Lambda$ is ruled out in this model.  A
laboratory detection of $(\dot{\alpha}/\alpha)_0$ may be possible in
the near future.

To satisfy the quasar, meteorite, and Oklo bounds on $\Delta
\alpha/\alpha$, the mass of the scalar field has to be on the order of
0.5--0.6 $H_\Lambda$.  It is difficult to satisfy both the
Oklo/meteorite and quasar bounds in theories where the variation of
$\alpha$ derives from the evolution of a scalar field; the scalar
field in the model studied here must be near an extremum near
$t_{Oklo}$ and $t_0$.

The key insight of this model, as well as other models of variable
$\alpha$, is that variation of $\alpha$ provides a window into the
parameters of the underlying theory that unifies gravity and the
standard model of particle physics.

\section*{Acknowledgment}

I would like to thank Thibault Damour for valuable comments.

\newpage

\begin{table}[h]
\center{
\begin{tabular}{|l|c|c|c|c|c|c|} \hline 
$n$ & $\zeta$ & $(\Delta \alpha/\alpha)_{BBN}$ & $(\Delta \alpha/\alpha)_{CMB}$
	& $(\Delta \alpha/\alpha)_{meteor}$ & $(\Delta \alpha/\alpha)_{Oklo}$
	& $(\dot{\alpha}/\alpha)_0$ \\ \hline
0	& $-2.0 \times 10^{-6}$ 
	& $-6.7 \times 10^{-5}$ 
	& $-5.3 \times 10^{-5}$ 
	& $-6.8 \times 10^{-7}$ 
	& $-4.8 \times 10^{-7}$ 
	& $ 2.4 \times 10^{-16}$ 
\\ \hline
0.3	& $-3.0 \times 10^{-6}$ 
	& $-9.6 \times 10^{-5}$ 
	& $-7.6 \times 10^{-5}$ 
	& $-1.4 \times 10^{-7}$ 
	& $-3.2 \times 10^{-9}$ 
	& $-6.8 \times 10^{-17}$ 
\\ \hline
0.31	& $-3.0 \times 10^{-6}$ 
	& $-9.8 \times 10^{-5}$ 
	& $-7.7 \times 10^{-5}$ 
	& $-1.1 \times 10^{-7}$ 
	& $ 2.0 \times 10^{-8}$ 
	& $-8.3 \times 10^{-17}$ 
\\ \hline
1	& $ 6.1 \times 10^{-6}$ 
	& $ 1.8 \times 10^{-4}$ 
	& $ 1.4 \times 10^{-4}$ 
	& $-3.7 \times 10^{-6}$ 
	& $-3.2 \times 10^{-6}$ 
	& $ 2.0 \times 10^{-15}$ 
\\ \hline
2	& $ 1.9 \times 10^{-6}$ 
	& $ 5.0 \times 10^{-5}$ 
	& $ 3.8 \times 10^{-5}$ 
	& $-2.7 \times 10^{-6}$ 
	& $-2.2 \times 10^{-6}$ 
	& $ 1.3 \times 10^{-15}$ 
\\ \hline
3	& $ 1.0 \times 10^{-6}$ 
	& $ 2.4 \times 10^{-5}$ 
	& $ 1.7 \times 10^{-5}$ 
	& $-2.2 \times 10^{-6}$ 
	& $-1.8 \times 10^{-6}$ 
	& $ 1.1 \times 10^{-15}$ 
\\ \hline
4	& $ 7.0 \times 10^{-7}$ 
	& $ 1.5 \times 10^{-5}$ 
	& $ 1.0 \times 10^{-5}$ 
	& $-2.0 \times 10^{-6}$ 
	& $-1.6 \times 10^{-6}$ 
	& $ 9.3 \times 10^{-16}$ 
\\ \hline
6	& $ 4.6 \times 10^{-7}$ 
	& $ 7.6 \times 10^{-6}$ 
	& $ 4.4 \times 10^{-6}$ 
	& $-1.8 \times 10^{-6}$ 
	& $-1.4 \times 10^{-6}$ 
	& $ 8.0 \times 10^{-16}$ 
\\ \hline
12	& $ 2.6 \times 10^{-7}$ 
	& $ 1.5 \times 10^{-6}$ 
	& $-2.8 \times 10^{-7}$ 
	& $-1.4 \times 10^{-6}$ 
	& $-1.1 \times 10^{-6}$ 
	& $ 5.7 \times 10^{-16}$ 
\\ \hline
\end{tabular}
\caption{Values of $\zeta$, BBN, CMB, meteorite, and Oklo 
$\Delta \alpha/\alpha$, and $(\dot{\alpha}/\alpha)_0$ in yr$^{-1}$ vs.\ 
scalar field mass squared $n$.}}
\end{table}

\newpage

\end{document}